\documentclass[11pt]{article}
\usepackage{amsfonts}
\usepackage{amssymb,latexsym}
\usepackage{amsmath,amscd}
\usepackage{theorem}
\usepackage[all]{xy}

\newtheorem{lemma} {Lemma} [section]
\newtheorem{proposition} [lemma] {Proposition}

\theorembodyfont{\rm}
\newtheorem{example}[lemma] {Example}
\newtheorem{remark}[lemma]{Remark}
\newenvironment{proof}{{\sc Proof:}}{{\hspace*{\fill} $\square$\\}}

\input{tcilatex}

\begin{document}

\title{\textbf{A Generalized Montgomery Phase Formula for Rotating Self
Deforming Bodies}}
\author{Alejandro Cabrera\thanks{Corresponding author. E-mail: \texttt{cabrera@mate.unlp.edu.ar}
\newline \mbox{~~~~~} Tel.:+54-221-4229850 int 105 Fax:+54-221-4245875.
} \\
%EndAName
Departamento de Matem\'{a}ticas \\
Universidad de La Plata\\
Calle 50 esq 115 (1900)\\
La Plata, Argentina}

\maketitle

\begin{abstract}
We study the motion of self deforming bodies with non zero angular
momentum when the changing shape is known as a function of time.
The conserved angular momentum with respect to the center of mass,
when seen from a rotating frame, describes a curve on a sphere as
it happens for the rigid body motion, though obeying a more
complicated non-autonomous equation. We observe that if, after
time $\Delta T$, this curve is simple and closed, the deforming
body
%TCIMACRO{\U{b4}}%
%BeginExpansion
\'{}%
%EndExpansion
s orientation in space is fully characterized by an angle or phase $\theta
_{M}$. We also give a reconstruction formula for this angle which
generalizes R. Montgomery%
%TCIMACRO{\U{b4}}%
%BeginExpansion
\'{}%
%EndExpansion
s well known formula for the rigid body phase. Finally, we apply
these techniques to obtain analytical results on the motion of
deforming bodies in some concrete examples.
\end{abstract}

MSC2000 Subject Classification Numbers: 53Z05 (Primary), 70F99
74A05 93B29 (secondary)

JGP SC: Classical mechanics, real and complex differential
geometry


\tableofcontents

\section{Introduction}

\subsection{Background}

We are going to study the problem of describing the motion of a rotating
body whose shape is changing with time in a known controlled fashion. A
particular case of this problem is the one in which the body%
%TCIMACRO{\U{b4}}%
%BeginExpansion
\'{}%
%EndExpansion
s shape is constant in time, i.e. a \emph{rigid body}.

As well known, a free \emph{rigid body} rotates about its center of mass in
a rather complicated way, depending on how its mass is distributed in space.
This distribution is represented by the corresponding \emph{inertia tensor}
and the motion is such that the \emph{(spatial) angular momentum} with
respect to the \emph{center of mass }is a conserved quantity.

Analytically, the orientation of the body with respect to an inertial
reference frame can be obtained by, first, solving \emph{Euler equations}
for the (body) angular momentum and, finally, \emph{reconstructing} the
desired curve in the space of rotations from the momentum one. A beautiful
result by R. Montgomery (\cite{Montph}) states that, when the momentum curve
completes a period, the orientation of the body in space is the initial one
up to a rotation in a certain angle about the (conserved) angular momentum
direction. Moreover, he derived a \emph{reconstruction formula }(see \cite%
{MMR}) for that angle, usually called the \emph{rigid body} \emph{phase,}
which involves a \emph{geometrical} (an holonomy) and a \emph{dynamical}
(energy and period values) contribution.

Now, when a rotating body is \emph{free}, but not rigid because
its shape changes with time in a prescribed fashion, the way in
which the mass is distributed in space is thus also changing with
time. How does such a body move? Or, since we know how its
\emph{shape} is changing: which is the rotation about the center
of mass induced by this changing mass distribution? For self
deforming bodies with \emph{zero angular momentum}, this question
was answered by Shapere and Wilczek in \cite{SW}. In that case,
the induced reorientation has a \emph{pure geometric nature}
because
it is described by a \emph{horizontal curve }with respect to the \emph{%
mechanical connection} in a $SO(3)$-principal fiber bundle over \emph{shape
space}\ (see \cite{SW}, \cite{Montgauge} and references therein)\emph{.}

Another related problem is that of finding the \emph{optimal} \emph{sequence
of deformations} in order to induce a \emph{given reorientation} of the
deforming body. This is an \emph{optimal control problem} which generalizes
the well known \emph{falling cat problem} (see \cite{Montgauge}). The
problem we want to analyze is, in a sense, \emph{the orthogonal} to the
above control problem: we \emph{know} the sequence of deformations and we
want to \emph{find} the induced reorientation.

\subsection{Main results}

In the present paper we shall focus on a case not covered in
\cite{SW}, i.e. the case in which a \emph{self deforming body}
rotates with \emph{non-zero} (conserved) angular momentum.

Our main result is an expression for an angle or \emph{phase} that
determines, at specific times, the \emph{exact orientation of a
spinning
self deforming body with non zero angular momentum}, generalizing R. Montgomery%
%TCIMACRO{\U{b4}}%
%BeginExpansion
\'{}%
%EndExpansion
s formula \cite{Montph}.

The examples that we shall be keeping in mind are the ones in which \emph{%
someone reaccommodates the furniture in a spacecraft} or \emph{an antenna
coming out from a satellite in orbit}.

Notice that in the above concrete examples, the body is acted by
external forces (e.g. gravity). Nevertheless, also note that for
\emph{small} objects like satellites in orbit the angular momentum
with respect to the center of mass is approximately conserved.
Within this approximation, the full motion can be described by two
sets of \emph{decoupled} eqations: the ones for the center of mass
(a \emph{central force problem}) and the ones we shall give below
for the rotation about the center of mass (a \emph{self deforming
body problem}).

The total reorientation of a self deforming body has two contributions: the
one induced from the change in its shape (of \emph{geometric nature} \cite%
{SW}) and the one we shall study, that follows from having non vanishing
angular momentum (of \emph{dynamical nature }as for a rigid body).

The class of deforming bodies we shall consider is the one that
will be refered to as \emph{self deforming bodies}. This bodies
are defined by a pure \emph{kinematical constraint} and a
\emph{dynamical hypothesis} described in section
\ref{sec:Semirigidbody}. In sec. \ref{sec:Eqs. motion}, we shall
derive the corresponding set of (second order) \emph{non
autonomous equations of motion} for the unknown rotation about the
center of mass. These follow from the \emph{conservation of the
angular momentum} measured from a reference system having its
origin at the center of mass and axes parallel to those of an
inertial one for all time. We will refer to it as the
\emph{spatial angular momentum}.

Also in \ref{sec:Eqs. motion}, we shall observe that, as in the rigid body
problem, the desired induced rotation can be \emph{reconstructed} from a
solution of the associated \emph{body angular momentum} (first order, non
autonomous)\ equations. This is the angular momentum as seen from a
reference frame which is \emph{rotating} \emph{with} the deforming body (see
\cite{SW}). At this point, we can re-state our main result: when, after some
time, the body angular momentum solution returns to its initial value, the
reconstructed rotation curve returns to its initial value up to a rotation
about the (conserved) spatial angular momentum direction; moreover, in
section \ref{sec:phases for semi rigid body} we show that the angle of this
rotation or \emph{self deforming body} \emph{phase} can be expressed (mod. $%
2\pi $) by the reconstruction formula $\left( \ref{Mont}\right) $ involving
a \emph{geometric} and \emph{dynamic} term. This result can be seen as a
straightforward generalization of Montgomery%
%TCIMACRO{\U{b4}}%
%BeginExpansion
\'{}%
%EndExpansion
s formula from the rigid body to the self deforming body motion.

This formula relates the body%
%TCIMACRO{\U{b4}}%
%BeginExpansion
\'{}%
%EndExpansion
s orientation with the (non conserved) \emph{energy integral} and the \emph{%
geometry} of the (non zero) body angular momentum solution curve. In the
zero angular momentum case of \cite{SW}, as the motion is of a pure
geometrical nature, the above phase becomes trivial.

As in the rigid body case, our formula can be applied when we have
a geometric description of an underlying simple closed body
angular momentum solution curve. Explicit time dependence of the
equations implies that, in general, \emph{energy is not conserved}
during the motion of the body. Also, as the equations for the body
angular momentum are non linear and have generic time-dependent
coefficients, solutions are hard to describe in the general case.

In view of this last observation, in sec. \ref{sec:Applications} we complete
this work by studying some particular classes of deformations. In each case,
we shall be able to derive analytical results on the motion of the
underlying deforming body by making simple dynamical estimates on the
geometry of the body angular momentum solutions and by thus applying the
generalized Montgomery formula.

\bigskip \bigskip

\noindent \textbf{Acknowledgements:} A.C. would like to thank Dr.
J. Solomin for stimulating discussions and suggestions. He would
also like to thank CONICET-Argentina for financial support.

\bigskip

\section{Physical setting}

\subsection{Deformable bodies}

\label{sec:Deformable bodies}Now, we review the setting presented in \cite%
{Montgauge} (see also \cite{SW}, \cite{LJ}) for deformable bodies.

Let us call $Q$ the \textbf{configurations space} of a system of $N-$%
particles or an extended body from the reference system $CM(t)$. Thus, $Q=%
\mathbb{R}^{3N-3}$ or $Q\equiv \{$smooth maps $q:B\subset \mathbb{R}%
^{3}\longrightarrow \mathbb{R}^{3}\ $s.t.$\ \int_{B}dvol\ \rho (x)\ q(x)=0\}$
for $B$ being a reference shape of the extended body. In both cases, the
usual action of $SO(3)$ on $\mathbb{R}^{3}$ gives rise to an action of $%
SO(3) $ on $Q.$ This action turns out to be free on
\begin{equation*}
Q_{0}=Q-Q_{1D}
\end{equation*}%
where $Q_{1D}$ is the set of points in $Q$ representing configurations in
which all the particles or the entire body is contained in a straight line.
Hence,%
\begin{equation*}
Q_{0}\overset{\pi }{\rightarrow }Q_{0}/SO(3)
\end{equation*}%
defines a \textbf{principal fiber bundle}, whose base $B=Q_{0}/SO(3)$ is
usually called the \textbf{shape space}.

In both particle system and extended body cases, the manifold $Q_{0}$ (and
also $Q)$ has a \textbf{Riemannian structure} induced by the usual scalar
product of $\mathbb{R}^{3}.$ So there is a natural \emph{principal connection%
} on the bundle $Q_{0}\overset{\pi }{\rightarrow }Q_{0}/SO(3)$
defined by choosing as the horizontal subspaces the orthogonal
complement to the vertical subspaces with respect to this metric.
This is usually called the
\textbf{mechanical connection}\emph{\ }on the bundle $Q_{0}\overset{\pi }{%
\rightarrow }Q_{0}/SO(3)$.

\begin{description}
\item
\begin{itemize}
\item[\textbf{Notation:}] \label{notation} From now on,

\item $S$ will denote a given \textbf{inertial reference frame},

\item $CM(t)$ will denote the reference frame with \textbf{origin at the
center of mass} of the body $r_{CM}(t)$ for each $t$ and \textbf{axes
parallel to those of} $S,$

\item $\widetilde{CM}(t)$ will denote any reference frame with \textbf{%
origin at the center of mass} of the body, with (possibly) \textbf{rotating
axes with respect to those of} $CM(t)$.
\end{itemize}
\end{description}

\begin{remark}
\emph{(Reference systems)} Note that a point $q_{0}\in $\textbf{\ }$Q_{0}$
over a shape $\pi (q_{0})=b_{0}\in Q_{0}/SO(3)$ gives the configuration of a
body with shape represented by $b_{0}$ as seen from the reference system $%
CM(t).$ Another point $\tilde{q}_{0}$ s.t. $\pi (\tilde{q}_{0})=\pi (q_{0})$
then represents the configuration, as seen from $CM(t),$ of a body with the
same shape but, now, rotated with respect to the one represented by $q_{0}.$
We can also interpret $\tilde{q}_{0}$ as describing the \emph{same body} but
as seen from a rotated reference system $\widetilde{CM}(t)$. This last
interpretation of the different points of a fiber $\pi ^{-1}(b_{0})$ is the
one that we shall keep in mind for the rest of the paper. See also the
discussion in ref. \cite{SW}.
\end{remark}

\subsection{Self deforming body hypothesis}

\label{sec:Semirigidbody}\label{def:self def body}Let us call $\tilde{r}%
_{io}(t)$ the position of the $i-$th particle with respect to $\widetilde{S}%
(t)$ at time $t.$ Then, for each time $t,$ there exist a global rotation $%
R(t)\in SO(3)$ and a translation $T(t)\in \mathbb{R}^{3}\ $such that the
position with respect to the inertial reference frame $S$ is%
\begin{equation}
r_{i}=R(t)\tilde{r}_{io}(t)+T(t).  \label{ec1}
\end{equation}

A \textbf{self deforming body} is defined to be a system of particles or an
extended object satisfying:

\begin{itemize}
\item[i)] \textbf{Kinematics:\ }There exist a \emph{reference frame} $%
\widetilde{S}(t),$ not necessarily inertial, from which we know $\tilde{r}%
_{io}(t)$ or, equivalently, a \emph{reference curve} $d_{0}(t)$ in $Q_{0}$.
Consequently, we also have a corresponding \emph{shape space curve} $\tilde{c%
}(t)=\pi (d_{0}(t)).$

\item[ii)] \textbf{Dynamics:\ }The constraint forces which act on the
particles in order to give this prescribed motions $\tilde{r}_{io}(t)$ are
\emph{internal forces }satisfying the \emph{strong action-reaction principle}%
. This means that all forces acting on the particle $i$ are caused by other
particles $j$%
%TCIMACRO{\U{b4}}%
%BeginExpansion
\'{}%
%EndExpansion
s and $F_{ij}^{int}=-F_{ji}^{int}$ with $F_{ij}^{int}$ parallel to the
vector $r_{ij}=r_{i}-r_{j}$.
\end{itemize}

Condition $(i)$ can be seen as a set of \emph{time dependent kinematical
constraints} generalizing the usual ones of rigidity: from $\widetilde{S}(t)$
we know how the body%
%TCIMACRO{\U{b4}}%
%BeginExpansion
\'{}%
%EndExpansion
s shape is changing (see also \cite{SW}).

\begin{example}
\emph{(Space-craft)}\textbf{\ }For the system being a space-craft, $%
\widetilde{S}(t)$ could be chosen as a frame fixed to some part of
the ship or an astronaut himself.
\end{example}

\begin{remark}
\emph{(Mechanical forces)}\textbf{\ }Notice that, although some forces do
not satisfy the strong action-reaction principle (for instance,
electro-magnetic forces), most of mechanical forces do.
\end{remark}

\begin{remark}
\emph{(Center of mass reference) }We can always take $\widetilde{S}(t)=%
\widetilde{CM}(t)$ (recall our notation \ref{notation}) having its
origin at the center of mass at all time. See also the discussion
at end of this section.
\end{remark}

\begin{remark}
\emph{(Non conservation of energy) }Note that with these kind of time
dependent constraints, the \emph{energy is not conserved} in general because
the deformation is implemented by time dependent constraint forces.
\end{remark}

The\textbf{\ self deforming body problem} is to find a curve{\Large \ }$R(t)$
in $SO(3)$ such that for%
\begin{equation}
\fbox{$c(t)=R(t)\cdot d_{0}(t)$}  \label{eqc(t)}
\end{equation}%
in $Q_{0}$ the \textbf{spatial angular momentum with respect to the }$c.m.$
\textbf{is conserved} (see below)\textbf{. }This can also be seen as a
\textbf{reconstruction problem }(see \cite{MMR}) for the rotation $R(t)$
from the given$\ \tilde{c}(t).$

\label{rmk:measurmentd}We end this section with some remarks on the meaning
and the measurement of $d_{0}(t)$\emph{. }First, we would like to stress
that the reference curve $d_{0}(t)$ is a \emph{natural physical input} for
the problem. To illustrate this fact, let us suppose that we want to
describe the motion of a space-craft or satellite when someone is reordering
the furniture inside of it, or when an antenna is coming out from this
satellite. Before launching, in the lab., an engineer can attach the
satellite to the floor and perform exactly the same deformation that will
occur in space. The body does not rotate because it is attached, but the
position of all its parts can be measured as a function of time $t$ from a
lab. reference frame. Then, the position of the center of mass can be
established for all $t$ and, consequently, the position of every part of the
body from a reference system $\widetilde{CM}(t)$ fixed to the center of mass
can be known for each $t.$

This provides us with a curve $d_{0}(t)$ as desired: when the satellite is
in orbit, the same deformation will occur yielding that $d_{0}(t)$ projects
onto the same curve in shape space as the \emph{physical curve} $c(t).$
Notice that, as the body can freely rotate about its center of mass, the
position with respect to $CM(t),$ represented by $c(t),$ will differ, in
general, by a rotation from the one described by $d_{0}(t)$ for each $t.$
This rotation is precisely the solution $R(t)$ of the \textbf{self deforming
body problem}.

\begin{example}
\emph{(Rigid body)} Note that the rigid body is a special case of the self
deforming body: take $\tilde{r}_{io}(t)$ constant for all time. More
generally, $d_{0}(t)$ must be contained in the fiber over the point
representing the constant shape of the rigid body for all $t$.
\end{example}

\subsection{Equations of motion}

\label{sec:Eqs. motion}The equations for $R(t)$, according to our definition
of the self deforming body, can be derived from the \textbf{conservation of
the angular momentum relative to the center of mass}\emph{\ }%
\begin{equation*}
\overset{\bullet }{L_{CM}}=0.
\end{equation*}%
This means that the rotation must be such that, from a frame $CM(t)\ $this
quantity is conserved even though things are moving internally in the system.

Let us recall the well known quantities: for $R(t)\in SO(3)\ $and $%
d(t)\equiv \{r_{i}(t)\}\in Q_{0},\ $

\begin{itemize}
\item \textbf{body angular velocity}\emph{\ }$\omega _{B}^{R(t)}\simeq R^{-1}%
\dot{R}$ is defined, as usual, by $\omega _{B}^{R(t)}\times v=R^{-1}\overset{%
\cdot }{R}v$ for all $v\in \mathbb{R}^{3}$. We shall denote $\Psi
:(so(3),[,])\longrightarrow (\mathbb{R}^{3},\times )$ the usual Lie algebra
isomorphism (see for ex. \cite{MR}).;

\item (locked)\ \textbf{Inertia tensor}:\ $I:Q_{0}\rightarrow
S_{>0}^{3\times 3}=\{3\times 3$ real symmetric positive definited matrices$%
\} $,\ $v\cdot I(\{r_{i}\})w=\underset{i}{\sum }m_{i}(v\times r_{i})\cdot
(w\times r_{i})$;

\item \textbf{Angular momentum} (with respect to a rotated frame with origin
at the center of mass): $L:TQ_{0}\rightarrow
%TCIMACRO{\U{211d} }%
%BeginExpansion
\mathbb{R}
%EndExpansion
^{3},$ $L(\{r_{i},\dot{r}_{i}\})=\underset{i}{\sum }m_{i}\ r_{i}\times \dot{r%
}_{i},$ satisfying%
\begin{equation}
\fbox{ $L(\frac{d}{dt}(R(t)d(t)))=R(t)\ I(d(t))\omega _{B}^{R(t)}+R(t)\ L(%
\frac{d}{dt}(d(t)))$};  \label{eq:angmom}
\end{equation}%
This gives the \textbf{momentum map }for the $SO(3)$ action on $TQ_{0}$ (see
the details in \cite{MMR}, \cite{Montgauge});

\item \textbf{Kinetic energy}: $T:TQ_{0}\rightarrow
%TCIMACRO{\U{211d} }%
%BeginExpansion
\mathbb{R}
%EndExpansion
,\ T(\{r_{i},\dot{r}_{i}\})=\underset{i}{\sum }m_{i}\ \dot{r}_{i}^{2},$ for
which
\end{itemize}

\begin{equation}
T(\frac{d}{dt}(R(t)d(t)))=\frac{1}{2}\omega _{B}^{R(t)}\cdot I(d(t))\omega
_{B}^{R(t)}+L(\frac{d}{dt}(d(t)))\cdot \omega _{B}^{R(t)}+T(\frac{d}{dt}%
(d(t))).  \label{Kinen}
\end{equation}%
$.$

For the \emph{physical }curve $c(t)$ in $Q_{0},$ the following quantity must
be conserved:
\begin{equation*}
L_{CM}=L(\frac{d}{dt}c(t))=L(\frac{d}{dt}(R(t)d(t)))=R(t)\ I(d_{0}(t))\omega
_{B}^{R(t)}+R(t)\ L(\frac{d}{dt}(d_{0}(t)))
\end{equation*}%
Here $I(d_{0}(t))$ is interpreted as the \emph{inertia tensor measured from
the reference frame} $\tilde{S}(t)=\widetilde{CM}(t)$ and we shall call%
\begin{equation*}
L_{o}(t):=L(\frac{d}{dt}(d_{0}(t)))=\underset{i}{\sum }m_{i}\widetilde{r}%
_{io}(t)\times \overset{\bullet }{\widetilde{r}_{io}}(t)
\end{equation*}%
the \emph{internal }(or \emph{apparent} \cite{SW}) \emph{angular momentum}
with respect to $\widetilde{CM}(t).$\bigskip

The (\emph{time-dependent, second order}) \textbf{equations of motion} for $%
R(t)$ thus read%
\begin{eqnarray}
\frac{d}{dt}L(R(t)d_{0}(t)) &=&0  \label{ec3} \\
I(d_{0}(t))\overset{\bullet }{\omega _{B}} &=&I(d_{0}(t))\omega _{B}\times
\omega _{B}+L_{o}(t)\times \omega _{B}-\frac{d}{dt}(I(d_{0}(t)))\ \omega
_{B}-\frac{d}{dt}L_{o}(t)  \notag
\end{eqnarray}%
when we express them in terms of the body angular velocity $\omega _{B}.$

\bigskip The \textbf{reconstruction equations }for $R(t)$, once we solved
the previous one for $\omega _{B},$ are%
\begin{equation}
\overset{\cdot }{R}=R\ \hat{\omega}_{B}
\end{equation}%
where $\hat{\omega}_{B}=\Psi ^{-1}(\omega _{B})$ with $\Psi
:(so(3),[,])\longrightarrow (\mathbb{R}^{3},\times )$ the usual Lie algebra
isomorphism (see \cite{MR}). The \textbf{initial value} $R(t_{1})$ must be
such that $R(t_{1})d_{0}(t_{1})=c(t_{1})$ coincides with the initial value
of the problem.

\begin{example}
\emph{(Rigid body)}\textbf{\ }For the \emph{rigid body}, recall that $%
d_{0}(t)$ must be contained on the fiber over a point in shape space. We can
then choose the $d_{0}(t)\ $(equiv. $\widetilde{r}_{io}(t))$ to be constant
for all $t,$ so $I(d_{0}(t))=I$ is constant in time and $L_{o}=0.$ In this
case, we recover \textbf{Euler equations}:%
\begin{equation*}
I\overset{\bullet }{\omega _{B}}=I\omega _{B}\times \omega _{B}
\end{equation*}%
as expected.
\end{example}

Also in analogy with the rigid body problem, as $L(\frac{d}{dt}c(t))\in
\mathbb{R}^{3}$ is conserved during the time evolution, if we define

\begin{equation}
\bigskip \Pi (t)=I(d_{0}(t))\omega _{B}^{R(t)}+L(\frac{d}{dt}d_{0})
\label{Pi}
\end{equation}%
we then have that $L(\frac{d}{dt}c(t))=R(t)\bigskip \Pi (t)$ and, hence, its
$\mathbb{R}^{3}-$norm $\left\Vert L(\frac{d}{dt}c(t))\right\Vert =\left\Vert
\Pi (t)\right\Vert $

is constant for all $t.$ The quantity $\Pi (t)$ represents the \emph{angular
momentum measured from the reference frame} $\tilde{S}(t)$ or \textbf{body
angular momentum.}

\begin{remark}
\emph{(Recovering the angular velocity) }Since $I(d_{0}(t))$ is invertible
for all $t,$ we can recover at every time $t$ the angular velocity $\omega
_{B}^{R(t)}$ from $\Pi (t)\in \mathbb{R}^{3}:$
\end{remark}

\begin{equation}
\omega _{B}^{R(t)}=I^{-1}(d_{0}(t))(\Pi (t)-L(\frac{d}{dt}d_{0})),\ \
\forall t.  \label{ecvelpi}
\end{equation}

The corresponding \emph{non-autonomous differential equation } for $\Pi
(t)\in \mathbb{R}^{3}$ is%
\begin{gather}
\fbox{$\dot{\Pi}=$ $\Pi \times (I^{-1}(d_{0}(t))(\Pi -L(\frac{d}{dt}%
d_{0}(t))))$}  \label{ECsphere} \\
\Pi (t_{1})=R^{-1}(t_{1})L_{CM}  \notag
\end{gather}%
whose solutions lie entirely on the \emph{sphere }$S_{L}^{2}$ $\subseteq
\mathbb{R}^{3}$ \emph{of radius} $\left\Vert L(\frac{d}{dt}c(t))\right\Vert
=\left\Vert \pi \right\Vert .$

Using $\left( \ref{ecvelpi}\right) $, the reconstruction equations for $%
R(t)\ $become%
\begin{equation}
\overset{\cdot }{R}=R\ \Psi ^{-1}(I^{-1}(d_{0}(t))(\Pi (t)-L(\frac{d}{dt}%
d_{0})))  \label{reconstruc}
\end{equation}%
or, equivalently, if we set $R(t_{1})=id$ for simplicity
\begin{equation*}
R(t)=Texp\int_{t_{1}}^{t_{2}}ds\ \Psi ^{-1}(I^{-1}(d_{0}(s))(\Pi (s)-L(\frac{%
d}{dt}d_{0}(s))))
\end{equation*}%
where $T$ stands for the time ordered integral (see also \cite{SW}).

\begin{remark}
\emph{(Non integrability) }In general, as noted before, the explicit time
dependence of the self deforming body tells us that \textbf{energy is not
conserved} and consequently, we cannot reduce the dimension of the problem
any further.
\end{remark}

\subsubsection{Gauge freedom}

By definition, we are given a curve $d_{0}(t)$ in the configuration space $%
Q_{0}$, but we might want to work with another curve $\tilde{d}_{0}(t)$
defining an equivalent self deforming body problem, i.e. $\pi (\tilde{d}%
_{0}(t))=\pi (d_{0}(t))=$ $\tilde{c}(t)$ $\in Q_{0}/SO(3)$. This is
equivalent to consider the self deforming body to be described from a new
reference frame $\widetilde{\widetilde{S}}(t)$ having the same origin and
rotating, in a certain known way, with respect to the initial one $\tilde{S}%
(t)$ from which the motion represented by $d_{0}(t)$ was originally
described.

\begin{remark}
\emph{(Gauge transformations)\ }This freedom in choosing the initial
orientation curve $d_{0}(t)$ can be seen as \emph{gauge freedom}.
Correspondingly, the change $d_{0}(t)\rightsquigarrow $ $\tilde{d}_{0}(t)$
can be thought of as a \textbf{gauge transformation}. For more details on
this analogy, we refer the interested reader to \cite{Montgauge}, \cite{SW}
and references therein.
\end{remark}

Among all possible \emph{lifts} $d_{0}(t)$ of $\tilde{c}(t)$ we consider two:

\begin{itemize}
\item[a)] the \textbf{horizontal lift} with respect to the \emph{mechanical
connection }in the bundle $Q_{0}\longrightarrow Q_{0}/SO(3)$. This is
equivalent to the problem of finding a lift $\tilde{d}_{0}(t)$ such that $L(%
\frac{d}{dt}\tilde{d}_{0}(t))=0$ $\forall t$ (see also remark \ref{rmk:zero
ang mom}).

\item[b)] a lift $\tilde{d}_{0}(t)$ for which the \textbf{inertia tensor} $I(%
\tilde{d}_{0}(t))$ is \textbf{diagonal} for all $t$. This is equivalent to
solve the problem of finding a lift of the base curve $I(d_{0}(t))$ along
the map
\begin{eqnarray*}
\mathfrak{A}\times SO(3) &\rightarrow &S_{>0}^{3\times 3} \\
(a,R) &\longmapsto &RaR^{-1}
\end{eqnarray*}%
where $\mathfrak{A}:=\{3\times 3$ diagonal positive definited matrices$\}.$
\end{itemize}

\begin{remark}
\label{rmk:zero ang mom}\emph{(Deformable bodies with zero angular
momentum) }In ref. \cite{SW}, it is shown that, given a shape
space curve, the motion of a self deforming body with zero angular
momentum is described by the corresponding horizontal lift as in
$(a)$ above. These computations also arise in the \emph{falling
cat problem} (see \cite{Montgauge}\cite{LJ}) and other interesting
problems (see references in \cite{SW}).
\end{remark}

\begin{remark}
\emph{(Simplifying the equations) }Choosing a different $d_{0}(t)$ changes
the time-dependence of the coefficients of equation $\left( \ref{ECsphere}%
\right) $. Thus, an appropriate choice could turn this equation into a \emph{%
simpler equivalent one}. For example, choosing the horizontal lift
implies that the equation has vanishing $L(d_{0}(t))$ term because
this is zero by construction. We also see that there is an obvious
simplification when choosing the lift keeping the inertia tensor
$I(d_{0}(t))$ diagonal. But observe that this two simplifications
\emph{cannot always be carried out at the same time,} since the
horizontal lift does not necessarily diagonalize the inertia
tensor in general.
\end{remark}

\bigskip

\section{Phases in the self deforming body motion}

\label{sec:phases for semi rigid body}

\subsection{Reconstruction}

For completeness, we now describe two types of
\textbf{reconstruction phases }(\cite{MMR}) appearing in the
configuration space during the motion of the body. In the rest of
the paper, we shall focus only on the second (\emph{abelian})\
one.

\subsubsection{Reconstruction of $c(t)$ from $\tilde{c}(t)$ in the bundle $%
Q_{0}\protect\overset{\protect\pi }{\rightarrow }Q_{0}/SO(3):$}

Recall that, for each $t,$ both $d_{0}(t)$ and $c(t)$ belong to the fiber
over $\tilde{c}(t)\ $in shape space $Q_{0}/SO(3)$ (see section \ref%
{sec:Semirigidbody}). When the shape space curve is closed in $[t_{1},t_{2}]$%
, we can then follow the \textbf{standard procedure for reconstruction}\ (%
\cite{MMR}):\ choose $d_{0}(t)$ to be the horizontal lift with respect to
the \emph{mechanical connection} having $d_{0}(t_{1})=c(t_{1})$ as initial
value. Then, $d_{0}(t_{2})=R_{G}\ c(t_{1})$ with $R_{G}$ being the \emph{%
holonomy} of the base path $\tilde{c}(t)$ measured from $c(t_{1})$ with
respect to this connection (see remark \ref{rmk:zero ang mom} and sec. \ref%
{sec:Deformable bodies}). This is often called the (\emph{non abelian})\emph{%
\ }\textbf{geometric phase}$.$\emph{\ }Finally, under these assumptions, the
reconstruction formula reads%
\begin{equation*}
c(t_{2})=\ R_{D}(t_{2})\ R_{G}(t_{1})\ c(t_{1})
\end{equation*}%
where $R_{D}(t_{2})$ is usually called the (non abelian) \textbf{dynamical
phase}\emph{.} This dynamical phase can be obtained by solving eq. $\left( %
\ref{ec3}\right) $ with the initial value $R_{D}(t_{1})=Id$ and with the
above \emph{horizontal} choice of $d_{0}(t),$ i.e. with $L(\dot{d}_{0})=0$.
For details on general reconstruction see \cite{MMR}. The interested reader
can find details about this reconstruction for a deforming body motion with
\emph{zero angular momentum} in \cite{SW}. For a study of phases in the $N=3$
body problem, we refer the interested reader to \cite{Mont3B}.

\bigskip

\subsubsection{Reconstruction of $R(t)$ from $\Pi (t)$ in the bundle $%
SO(3)\longrightarrow S_{L}^{2}:$}

Recall that, in general, the unknown rotation $R(t)$ in eq. $\left( \ref%
{eqc(t)}\right) $ can be reconstructed via $\left( \ref{reconstruc}\right) $
once we have solved the equation $\left( \ref{ECsphere}\right) $ on the
sphere. An interesting special case is when this solution $\Pi (t)$ is
closed in the interval $[t_{1},t_{2}],$ that is when%
\begin{equation*}
\Pi (t_{1})=\Pi (t_{2}).
\end{equation*}%
In this case, there is a unique angle $\theta _{M}$ naturally associated to
this solution and to the initial condition $R(t_{1})$ such that%
\begin{equation*}
R(t_{2})=exp(\theta _{M}\ \frac{\hat{L}}{\left\Vert L\right\Vert })\
R(t_{1}),
\end{equation*}%
yielding%
\begin{equation*}
\fbox{ $c(t_{2})=[exp(\theta _{M}\ \frac{\hat{L}}{\left\Vert L\right\Vert }%
)\ R(t_{1})]\ d_{0}(t_{2})$}
\end{equation*}%
where $\hat{L}=\Psi ^{-1}(L)\in so(3).$ We see that $\theta _{M}$ defines an%
\emph{\ }\textbf{abelian reconstruction phase} associated to the initial
data $R(t_{1})$ (coming from $c(t_{1})$). This phase appears when
reconstructing $R(t)\ $from $\Pi (t)$ in a $U(1)$-principal bundle $%
SO(3)\longrightarrow S_{L}^{2}$ that we shall describe in the next section.

\begin{remark}
\emph{(Interpretation of }$\theta _{M}\emph{)}$ Recall that $R(t)$ takes the
reference frame $\widetilde{CM}(t)$ to $CM(t).$ This implies that, at time $%
t_{2}$ as above, the orientation of the body, as seen from
$CM(t)$, is precisely obtained by rotating the known configuration
$d_{0}(t_{2})$ about the conserved angular momentum direction
($L_{CM}$) in the angle $\theta _{M}.$ So this phase \emph{fully
characterizes} the position of the deforming body in space at
specific times (i.e. $t_{2}$).
\end{remark}

In the rest of the paper, we shall focus on the latter
reconstruction procedure. Note that, as the second phase is
abelian, it is more likely to have simpler closed expressions for
its reconstruction.

Finally, we note that the most geometrically interesting situation is that
in which both, the solution $\Pi (t)$ to $\left( \ref{ECsphere}\right) $ and
the shape space base curve $\tilde{c}(t)$ in $Q_{0}/SO(3),$ are closed in
the same interval $[t_{1},t_{2}],$ i.e.,%
\begin{eqnarray*}
\Pi (t_{1}) &=&\Pi (t_{2}) \\
\tilde{c}(t_{1}) &=&\tilde{c}(t_{2}).
\end{eqnarray*}%
When this conditions hold, there is a geometrically defined phase in the
bundle $Q_{0}\overset{\pi }{\rightarrow }Q_{0}/SO(3)$
\begin{equation*}
c(t_{2})=\Delta R\cdot c(t_{1})
\end{equation*}%
independent of the choice of $d_{0}(t)$ (it only depends on the initial
value $c(t_{1})$) and having the following expression:%
\begin{equation*}
\fbox{$\Delta R=exp(\theta _{M}\ \frac{\hat{L}}{\left\Vert L\right\Vert })\
R(t_{1})\Delta R_{0}R^{-1}(t_{1})$},
\end{equation*}%
where $\Delta R_{0}=R_{0}(t_{2})R_{0}^{-1}(t_{1})$ and $R(t_{1})$ are fixed
by the initial condition $c(t_{1})$ and the angle $\theta _{M}$ is again
given by the \emph{generalized Montgomery formula} presented in the next
section.

\subsection{Generalized Montgomery formula}

In this subsection, we give a \emph{phase formula} for the reconstruction of
the rotation $R(t)$ from a closed solution curve $\Pi (t)$ of the equation $%
\left( \ref{ECsphere}\right) $. This formula generalizes the well
known one given by R. Montgomery in \cite{Montph} for the rigid
body phase. For the proofs, we shall use some differential
geometric results that we review below.

\subsubsection{Preliminaries}

Recall the diagram (see \cite{MMR})
\begin{gather*}
so_{-}^{\ast }(3)\overset{\pi }{\longleftarrow }T^{\ast }SO(3)\overset{%
L^{\ast }}{\simeq }SO(3)\times so^{\ast }(3)\overset{J}{\longrightarrow }%
so_{-}^{\ast }(3) \\
\xi \longleftarrow (R,\xi )\longrightarrow Ad_{R}^{\ast }\xi
\end{gather*}%
where: $so_{-}^{\ast }(3)$ denotes the Poisson manifold $so^{\ast }(3)$ with
its (minus) standard Poisson bracket; $\pi $ and $J$ are Poisson and
anti-Poisson maps respectively and $Ad_{R}^{\ast }:=(Ad_{R^{-1}})^{\ast }$
denotes the (left) coadjoint action of $SO(3)\ $on $so^{\ast }(3)$. Recall
(see e.g. \cite{MR}) that $J$ is the \emph{momentum map} associated to the
left symplectic action of $SO(3)$ on $T^{\ast }SO(3)$. The trivialization $%
T^{\ast }SO(3)\overset{L^{\ast }}{\simeq }SO(3)\times so^{\ast }(3)$ by left
translations is known as passing to \emph{body coordinates}.

If we fix an element $L\in so_{-}^{\ast }(3)\simeq so(3)\simeq \mathbb{R}%
^{3} $ (the isomorphisms are compatible with the corresponding Poisson
brackets), then we have that%
\begin{equation*}
\Psi (Ad_{R}^{\ast }\xi )=R\cdot \Psi (\xi )
\end{equation*}%
thus%
\begin{equation*}
\pi (J^{-1}(L))=S_{L}^{2}.
\end{equation*}%
\bigskip The sphere $S_{L}^{2}$ of radius $\left\Vert L\right\Vert $ defines
a \emph{symplectic leaf} in $so_{-}^{\ast }(3)\simeq so(3)\simeq \mathbb{R}%
^{3}\ $(see \cite{MR})$.$ Moreover, in this case we have that%
\begin{gather*}
J^{-1}(L)=\{(R,\Pi );\ R\cdot \Pi =L\}\simeq SO(3)\overset{\pi }{%
\longrightarrow }S_{L}^{2} \\
(R,R^{-1}L)\longmapsto R^{-1}L
\end{gather*}%
is a $U(1)$-principal fiber bundle (see \cite{MMR}).

Now, consider the inclusion $J^{-1}(L)\overset{i}{\hookrightarrow }%
SO(3)\times so^{\ast }(3)\overset{L^{\ast }}{\simeq }T^{\ast }SO(3)$ and the
$u(1)-$valued 1-form on $J^{-1}(L)$
\begin{equation}
A:=\frac{1}{\left\Vert L\right\Vert }i^{\ast }\Theta ^{L}  \label{connect}
\end{equation}%
where $\Theta ^{L}$ is the canonical left invariant 1-form on $T^{\ast
}SO(3) $ in \emph{body coordinates}. It can be seen that $A$ gives a \emph{%
principal connection} in the principal $U(1)-$bundle $J^{-1}(L)\overset{\pi }%
{\longrightarrow }S_{L}^{2}$ (\cite{MMR}). This connection 1-form satisfies%
\begin{equation*}
dA=-\frac{1}{\left\Vert L\right\Vert }i^{\ast }\omega ^{L}
\end{equation*}%
where $\omega ^{L}=-d\Theta ^{L}$ denotes the canonical symplectic 2-form on
$T^{\ast }SO(3)$ in body coordinates.\emph{\ }By the \emph{reduction theorem} (%
\cite{MW}, see also \cite{MR}),
\begin{equation*}
i^{\ast }\omega ^{L}=\pi ^{\ast }\omega _{\mu }
\end{equation*}%
with $\omega _{\mu }$ the reduced symplectic form on $S_{L}^{2}$. Finally,
if $dS$ denotes the standard area $2$-form on the sphere $S_{L}^{2}\subseteq
\mathbb{R}^{3},$ then (see \cite{MR})%
\begin{equation*}
\omega _{\mu }=-\frac{1}{\left\Vert L\right\Vert }dS.
\end{equation*}

\subsubsection{The formula}

With these geometrical background, the following can be easily proved:

\begin{proposition}
$R(t)$ is a solution of the (second order) equation of motion $\left( \ref%
{ec3}\right) $ iff $(R(t),\Pi (t))\in $ $J^{-1}(L)\subset T^{\ast }SO(3)$ is
an integral curve of the \emph{time dependent} \emph{vector field}
\begin{equation*}
X(R,\Pi ,t)=(R\ \Psi ^{-1}(I^{-1}(d_{0}(t))(\Pi -L(\dot{d}_{0}))),\ \Pi
\times (I^{-1}(d_{0}(t))(\Pi -L(\dot{d}_{0})))).
\end{equation*}
\end{proposition}

\begin{remark}
\emph{(Hamiltonization) }This result can be viewed as a time dependent\emph{%
\ hamiltonization} from $TQ_{0}$ to $T^{\ast }SO(3)$ using the momentum map $%
L$ and also a further \emph{reduction }to $J^{-1}(L)$ of the problem
equations of motion $\left( \ref{ec3}\right) $. See also similar comments
about reduction for the $3$-body problem phases in \cite{Mont3B}.
\end{remark}

Thus, reconstructing $R(t)$ from $\Pi (t)$ is the same as finding\emph{\ }a
curve $(R(t),\Pi (t))\in J^{-1}(L)$ in the $U(1)-$bundle\emph{\ }$%
J^{-1}(L)\simeq SO(3)\overset{\pi }{\longrightarrow }S_{L}^{2}$ as above
such that the projection to the base\emph{\ }$\Pi (t)\in S_{L}^{2}$ is a
solution of $\left( \ref{ECsphere}\right) $. Given $\Pi (t),$ we can apply
the usual procedure of \emph{reconstruction} (\cite{MMR}): choose $%
R_{0}(t)\in J^{-1}(L)$ in a \emph{natural geometric way} as the \textbf{%
horizontal lift }of $\Pi (t)\in S_{L}^{2}$ from the initial value $%
R_{0}(t_{1})=R(t_{1})$ with respect to the connection $A.$ Now, let $\theta
(t)\in U(1)$ be an angle to be determined by requiring the curve%
\begin{equation*}
exp(\theta (t)\frac{\hat{L}}{\left\Vert L\right\Vert })\cdot
(R_{0}(t),R_{0}^{-1}(t)L)=(exp(\theta (t)\frac{\hat{L}}{\left\Vert
L\right\Vert })R_{0}(t),R_{0}^{-1}(t)L)\in J^{-1}(L)\simeq SO(3)
\end{equation*}%
to be the desired integral curve of $X(R,\Pi ,t)$. In the above formula, $%
\hat{L}$ denotes $\Psi ^{-1}(L)\in so(3).$

It follows that $\theta (t)$ must satisfy the following equation%
\begin{eqnarray}
\left\Vert L\right\Vert \overset{\bullet }{\theta }(t)
&=&I^{-1}(d_{0}(t))\Pi (t)\cdot \Pi (t)-I^{-1}(d_{0}(t))L_{0}(t)\cdot \Pi (t)
\label{tita} \\
\theta (t_{1}) &=&0  \notag
\end{eqnarray}

Now, note that if $[t_{1},t_{2}]\subseteq \mathbb{R}$ is a closed interval,
and $\Pi :[t_{1},t_{2}]\rightarrow S_{L}^{2}$ is (any) continuous curve,
then its image $Im(\Pi )$ is a compact, hence closed, subset of the sphere $%
S_{L}^{2}.$ So its complement $Im(\Pi )^{C}$ is open and it exists a closed
disc $\bar{d}\subseteq Im(\Pi )^{C}.$ We then have $Im(\Pi )\subseteq $ $%
\bar{d}^{C}$ and we thus showed

\begin{lemma}
\label{Lemma1}The image $Im(\Pi )$ of a continuous map $\Pi
:[t_{1},t_{2}]\rightarrow S_{L}^{2}$ is entirely contained in an open disc $%
D\subseteq S_{L}^{2}.$
\end{lemma}

We can now state our main result:

\begin{proposition}
\emph{(\textbf{generalized Montgomery formula}): }Let $\Pi (t)$ be a
solution of $\left( \ref{ECsphere}\right) $ satisfying that $\Pi (t_{1})=\Pi
(t_{2})$ for some interval $[t_{1},t_{2}]$ and that the image of $\Pi
:[t_{1},t_{2}]\rightarrow S_{L}^{2}$ is a \emph{simple} closed curve (i.e. $%
Im(\Pi )$ homeomorphic to the circle $S^{1}$) then $R(t_{2})=exp(\theta
_{M}\ \frac{\hat{L}}{\left\Vert L\right\Vert })\ R(t_{1})$ and the angle $%
\theta _{M}$ is given (mod 2$\pi $) by the formula%
\begin{equation}
\fbox{$\theta _{M}=(\mp )\frac{area(\tilde{D})}{\left\Vert L\right\Vert ^{2}}%
+\frac{1}{\left\Vert L\right\Vert }\int\limits_{t_{1}}^{t_{2}}dt\
(I^{-1}(d_{0}(t))\Pi (t)-I^{-1}(d_{0}(t))L(\dot{d}_{0}))\cdot \Pi (t)$}
\label{Mont}
\end{equation}%
where $\tilde{D}$ is a surface in $S_{L}^{2}$ bounded by the image of $\Pi $%
. The $-$ (resp. $+)$ sign corresponds to the case in which the
solid angle defined by $\tilde{D}$ on the sphere, with its
time-oriented boundary $\Pi (t)$, is a \emph{positive} (resp.
\emph{negative}) \emph{signed solid angle.}
\end{proposition}

\begin{remark}
\emph{(Signed solid angles) }As usual, we are considering a solid angle in
the sphere to be positive or negative by applying the \textbf{right hand rule%
} to its oriented boundary (see \cite{Montph}). Also notice that, $mod.$ $%
2\pi ,$ the above formula keeps the same form (i.e., with the $-$ sign) if
we replace $\frac{area(\tilde{D})}{\left\Vert L\right\Vert ^{2}}$ by the
corresponding \emph{signed solid angle}.
\end{remark}

\begin{remark}
\emph{(Relation to the energy)}\textbf{\ }The integrand in the right hand
side of this formula can be expressed in terms of the total kinetic energy
(see eq. $\left( \ref{Kinen}\right) $):%
\begin{equation*}
(I^{-1}(d_{0})\Pi (t)-I^{-1}(d_{0})L(\dot{d}_{0}))\cdot \Pi (t)=
\end{equation*}%
\begin{equation*}
=2T(\frac{d}{dt}(Rd_{0}))-2T(\frac{d}{dt}d_{0})+I^{-1}(d_{0})L(\dot{d}%
_{0})\cdot L(\dot{d}_{0})-I^{-1}(d_{0})L(\dot{d}_{0})\cdot \Pi (t)
\end{equation*}
\end{remark}

\begin{proof}
By the above mentioned reconstruction procedure and since $U(1)$ is abelian,%
\begin{eqnarray*}
R(t_{2}) &=&exp(\theta _{D}\ \frac{\hat{L}}{\left\Vert L\right\Vert })\cdot
exp(\theta _{G}\ \frac{\hat{L}}{\left\Vert L\right\Vert })\cdot R(t_{1}) \\
&=&exp((\overset{\theta _{M}}{\overbrace{\theta _{D}+\theta _{G}}})\ \frac{%
\hat{L}}{\left\Vert L\right\Vert })\cdot R(t_{1}),
\end{eqnarray*}

where $\theta _{D}$ is the \emph{dynamical phase,} solution of eq. $\left( %
\ref{tita}\right) $ and $\theta _{G}$ the \emph{geometric phase}, given by
the \textbf{holonomy} of the base path $\Pi (t)$ with respect to the
connection $A$ and measured from $R(t_{1}).$ Thus the dynamical contribution
$\theta _{D}$ to $\theta _{M}$ is precisely the second term in the r.h.s. of
eq. $\left( \ref{Mont}\right) .$

Let us then show that the remaining term coincides with the geometric
contribution $\theta _{G}$.

By the hypothesis and Lemma \ref{Lemma1}, $Im(\Pi )$ is entirely contained
in a smooth disc $D$ in $S_{L}^{2}$. Being $D$ contractible, the restricted
principal $U(1)-$bundle $J^{-1}(L)\mid _{D}\longrightarrow D$ is \emph{%
trivial} and, then, we have a smooth section $s:D\rightarrow J^{-1}(L)$.
Once we have chosen the disk $D$ containing the curve $Im(\Pi ),$ the
existence of a surface $\tilde{D}\subseteq S_{L}^{2}$ whose boundary is $\Pi
(t)$ is obvious since $D$ is diffeomorphic to an open disk in $\mathbb{R}%
^{2} $ and $Im(\Pi )$ is homeomorphic to $S^{1}.$ Thus, mod 2$\pi $, we can
write (see \cite{MMR})%
\begin{eqnarray*}
\theta _{G} &=&-\int \int_{\tilde{D}}s^{\ast }(dA) \\
&=&-\frac{1}{\left\Vert L\right\Vert ^{2}}\int \int_{\tilde{D}}dS=-\frac{%
area(\tilde{D})}{\left\Vert L\right\Vert ^{2}}
\end{eqnarray*}%
when the solid angle defined by $\tilde{D}$ is positively oriented with
respect to the (time oriented) boundary curve $\Pi (t)$. The last two
equalities follow from the results reviewed in the previous section. Formula
$\left( \ref{Mont}\right) $ is then completed.
\end{proof}

\begin{example}
\emph{(Rigid body)} For the rigid body, the kinetic energy $T$ is conserved
and, as we observed previously, $d_{0}(t)$ can be taken as a point for all $%
t $. So $L(\frac{d}{dt}d_{0})=0$ and the inertia tensor $I(d_{0})=I$ is
constant. In this case, the periodic solutions of Euler equations bound
disks on the sphere, thus $\tilde{D}$ defines the usual signed solid angle
and the above formula becomes the well known reconstruction formula derived
by R. Montgomery (\cite{Montph}).
\end{example}

\section{Some Applications}

\label{sec:Applications}

\subsection{Solutions on the sphere}

We shall now describe some tools which can be used to study the geometry of
solutions of eq. $\left( \ref{ECsphere}\right) $ on the sphere $S_{L}^{2}.$
Focusing on some particular cases we will be able to use this
characterization of the solutions to yield analytical results on the motion
of self deforming bodies by calculating the associated generalized
Montgomery phase $\theta _{M}$.

\begin{itemize}
\item \textbf{Reconstruction of }$R(t)$\textbf{: }When the solution $\Pi (t)$
for some time interval $[t_{A,}t_{B}]$ is an open path, we noted before that
the rotation $R(t)$ can be expressed as $exp(\theta (t)\frac{\hat{L}}{%
\left\Vert L\right\Vert })\cdot (R_{0}(t),R_{0}^{-1}(t)L),$ with $%
(R_{0}(t),R_{0}^{-1}(t)L)$ the horizontal lift of the base path $\Pi (t)$
with respect to the connection $\left( \ref{connect}\right) $ and $\theta
(t) $ a solution of eq. $\left( \ref{tita}\right) $. When the solution $\Pi
(t)$ is a closed simple curve for a time interval $[t_{A,}t_{B}]$, we have a
well defined phase $\theta _{M}$ given by formula $\left( \ref{Mont}\right) $%
. So, given a solution $\Pi (t)$ in $[t_{1,}t_{2}],$ we can find a \emph{%
total phase }by adding phases corresponding to sub-time intervals $%
[t_{i,}t_{i+1}]$ for which the solution is a \emph{simple open arc} in $%
S_{L}^{2}$ or a \emph{closed simple curve }in $S_{L}^{2}.$ In the first
case, we have a phase defined by%
\begin{equation*}
R(t_{i+1})=exp(\theta (t_{i+1})\frac{\hat{L}}{\left\Vert L\right\Vert }%
)Par(R(t_{i}))
\end{equation*}%
with $Par:\pi ^{-1}(\Pi (t_{i}))\longrightarrow \pi ^{-1}(\Pi (t_{i+1}))$
the \emph{parallel transport} (see \cite{MMR})\ in the $U(1)-$principal
bundle $J^{-1}(L)\overset{\pi }{\longrightarrow }S_{L}^{2}$ of the initial
condition $R(t_{i}),$ and $\theta (t)$ the solution of $\left( \ref{tita}%
\right) $ with $\theta (t_{i})=0.$ In the second case, fixing the initial
value $R(t_{i}),$ the phase is defined by $R(t_{i+1})=exp(\theta _{M}\ \hat{L%
})\ R(t_{i})$ with $\theta _{M}$ given by formula $\left( \ref{Mont}\right) $%
.

\item \textbf{The Energy:} As we noted before, in general, the energy is
\textbf{not a conserved quantity} during the self deforming body motion.
Nevertheless, if we know the evolution of the kinetic energy $T(\frac{d}{dt}%
(Rd_{0}))$ with time, we will be able to determine a specific subset of $%
S_{L}^{2}$ in which the corresponding solution $\Pi (t)$ lies$.$ This fact
can be shown as follows: let us define for each time $t$%
\begin{eqnarray*}
E_{t} &:&S_{L}^{2}\longrightarrow \mathbb{R} \\
&:&\Pi \longmapsto \frac{1}{2}\Pi \cdot I^{-1}(d_{0}(t))\ \Pi .
\end{eqnarray*}%
Note that%
\begin{equation*}
E_{t}(\Pi (t))=T(\frac{d}{dt}(Rd_{0})(t))-T(\frac{d}{dt}d_{0}(t))+\frac{1}{2}%
L(\dot{d}_{0}(t))\cdot I^{-1}(d_{0}(t))L(\dot{d}_{0}(t))
\end{equation*}%
for $\Pi (t)$ a solution of $\left( \ref{ECsphere}\right) $. Hence, as $%
d_{0}(t)$ is given, $E_{t}(\Pi (t))$ is uniquely determined by the kinetic
energy $T(\frac{d}{dt}(Rd_{0})(t))$. In this case, the corresponding
solution $\Pi (t)$ on the sphere at time $t$ must lie in the set%
\begin{equation*}
E_{t}^{-1}(k(t))\cap S_{L}^{2}
\end{equation*}%
where
\begin{equation*}
k(t)=T(\frac{d}{dt}(Rd_{0})(t))-T(\frac{d}{dt}d_{0}(t))+\frac{1}{2}%
L(d_{0}(t))\cdot I^{-1}(d_{0}(t))L(d_{0}(t)).
\end{equation*}%
The level sets $E_{t}^{-1}(k(t))$ are (generally non centered and rotated)
\emph{ellipsoids} for each $k\geqslant 0$ and each $t.$ Also notice that,
for a fixed time $t_{i},$ the intersection $E_{t_{i}}^{-1}(k(t_{i}))\cap
S_{L}^{2}$ gives the set where the \emph{body angular momentum of a rigid
body with constant inertia tensor equal to} $I(d_{0}(t_{0}))$ and \emph{%
energy} $k(t_{1})$ would lie. Finally, the equation for the evolution of $%
E_{t}(\Pi (t))$ is%
\begin{equation*}
\frac{d}{dt}E_{t}(\Pi (t))=[\Pi (t)\times I^{-1}(d_{0}(t))\Pi (t)]\cdot
I^{-1}(d_{0}(t))L(d_{0}(t))+\frac{1}{2}\Pi (t)\cdot \frac{d}{dt}%
[I^{-1}(d_{0}(t))]\Pi (t)
\end{equation*}%
which is coupled to the equation $\left( \ref{ECsphere}\right) $ for $\Pi
(t).$

\item \textbf{The arc-length:} for a given closed time interval $%
[t_{1},t_{2}]$ we are going to find a bound for the length of $\Pi
([t_{1},t_{2}]).$ To that end, we note that%
\begin{eqnarray*}
\left\Vert \frac{d}{dt}\Pi (t)\right\Vert &=&\left\Vert \Pi \times
(I^{-1}(d_{0}(t))(\Pi -L(\dot{d}_{0}(t))))\right\Vert \\
&\leq &\left\Vert \Pi \right\Vert (\left\Vert I^{-1}(d_{0}(t))\Pi
\right\Vert +\left\Vert I^{-1}(d_{0}(t))L(\dot{d}_{0}(t)))\right\Vert )
\end{eqnarray*}%
and, if $\left\Vert I^{-1}(d_{0}(t))v\right\Vert \leq a^{-1}(t)\left\Vert
v\right\Vert $ for all $v\in \mathbb{R}^{3},$ $t\in \lbrack t_{1},t_{2}],$
then%
\begin{equation*}
\left\Vert \frac{d}{dt}\Pi (t)\right\Vert \leq \left\Vert \Pi \right\Vert
^{2}a^{-1}(t)+\left\Vert \Pi \right\Vert a^{-1}(t)\left\Vert L(\dot{d}%
_{0}(t)))\right\Vert .
\end{equation*}%
Since $\left\Vert \Pi \right\Vert =\left\Vert L\right\Vert =l$ is constant,
we then have that%
\begin{equation*}
lenght(\Pi ([t_{1},t_{2}]))\leq l\int_{t_{1}}^{t_{2}}a^{-1}(t)(l+\left\Vert
L(\dot{d}_{0}(t))\right\Vert )dt
\end{equation*}%
When $a^{-1}(t)$ is a very small function (compared to $\frac{1}{%
l(t_{1}-t_{2})}$), we can deduce that $\Pi ([t_{1},t_{2}])$ is contained in
a \emph{small patch} in $S_{L}^{2}.$
\end{itemize}

For general time dependent parameters $I^{-1}(d_{0}(t))$ and $L(\dot{d}%
_{0}(t))$ we cannot give a characterization of the solution $\Pi (t)$ of eq.
$\left( \ref{ECsphere}\right) $. So we shall focus on some specific cases to
illustrate how to handle concrete problems.

\subsubsection{Cases with $I(t)=diag(I_{1}(t),I_{2}(t),I_{3}(t))$ and $L(%
\dot{d}_{0}(t))=0$ for all $t$.}

Let us denote by $(1,2,3)$ the cartesian axes of $so^{\ast }(3)\simeq $ $%
\mathbb{R}^{3}$and, hence, $\Pi =(\Pi _{1},\Pi _{2},\Pi _{3})\in \mathbb{R}%
^{3}$. In this cases, the intersections of each axis with the sphere $%
S_{L}^{2}$ give a constant solution of $\left( \ref{ECsphere}\right) $,
because at that points $\Pi $ is parallel to $I^{-1}(t)\Pi $ and the r.h.s.
of eq. $\left( \ref{ECsphere}\right) $ vanishes.

The equation for the evolution of the energy becomes

\begin{equation*}
\frac{d}{dt}E_{t}(\Pi (t))=\frac{1}{2}\Pi (t)\cdot \frac{d}{dt}%
[I^{-1}(d_{0}(t))]\Pi (t)
\end{equation*}%
and the arc-length is bounded by
\begin{equation*}
lenght(\Pi ([t_{1},t_{2}]))\leq l^{2}\int_{t_{1}}^{t_{2}}a^{-1}(t)dt.
\end{equation*}

Now, we shall analyze further the special case
\begin{equation*}
\fbox{$I_{1}(t)<I_{2}(t)<I_{3}(t)$}
\end{equation*}%
for all $t\in \lbrack t_{1},t_{2}].$ Notice that this is the case (up to
renumbering the $I_{i}%
%TCIMACRO{\U{b4}}%
%BeginExpansion
{\acute{}}%
%EndExpansion
s$) for small enough time intervals $[t_{1},t_{2}]$. Under this conditions,
the axes of the ellipsoids $E_{t}^{-1}(k(t))$ coincide with the cartesian
axes in $\mathbb{R}^{3}$ and the arc-length is thus bounded by%
\begin{equation*}
lenght(\Pi ([t_{1},t_{2}]))\leq l^{2}\int_{t_{1}}^{t_{2}}I_{1}^{-1}(t)dt.
\end{equation*}

Fixing the time $t,$ we have that through each point of $S_{L}^{2}$ passes a
solution of Euler equations (rigid body) with inertia tensor $%
diag(I_{1}(t),I_{2}(t),I_{3}(t))$. For each time we then have the
corresponding \textbf{homoclinic solutions} (see for ex. \cite{MMR})\emph{, }%
given by the intersection of $S_{L}^{2}$ with the ellipsoid of energy $k(t)=%
\frac{l^{2}}{I_{2}(t)}$.

Given a solution $\Pi (t)=(\Pi _{1}(t),\Pi _{2}(t),\Pi _{3}(t))$ of $\left( %
\ref{ECsphere}\right) $ for the interval $[t_{1},t_{2}]$ with initial value $%
\Pi (t_{1}),$ the function $f(t)=E_{t}(\Pi (t))$ reaches a maximum and a
minimum on $[t_{1},t_{2}]$, denoted $E_{max}$ and $E_{min}$ respectively.
The same happens with the value of the principal moments of inertia $%
I_{i}(t).$ The solution $\Pi (t)$ is then contained in a \textbf{connected
"crown like" region} $R$ which is the connected component of $\sqcup _{t\in
\lbrack t_{1},t_{2}]}E_{t}^{-1}([E_{min},E_{max}])\cap S_{L}^{2}$ which
contains the initial value $\Pi (t_{1}).$

We can now show the following results on the qualitative behavior of $\Pi
(t) $:

\begin{enumerate}
\item if \fbox{$E_{min}>\frac{l^{2}}{I_{2_{min}}}$} then $R$ is contained in
either the semi-space $\Pi _{1}>0$ or in $\Pi _{1}<0.$ In this case, $\Pi
(t) $ evolves in $S_{L}^{2}$ describing a trajectory that \emph{orbits
surrounding the cartesian axis} $1.$ More precisely, if the initial point
lies in the component with (say) $\Pi _{1}>0,$ then the solution will lie in
this component for all $t$ in $[t_{1},t_{2}]$. So if we consider spherical
coordinates $(\theta ,\varphi )$%
\begin{eqnarray*}
\Pi _{1} &=&l\ cos\theta \\
\Pi _{2} &=&l\ sin\theta \ cos\varphi \\
\Pi _{3} &=&l\ sin\theta \ sin\varphi
\end{eqnarray*}%
with $\theta \in \lbrack 0,\pi ],$ $\varphi \in \lbrack 0,2\pi ],$ it
follows that $\theta (t)<\frac{\pi }{2}$ for all $t$ in $[t_{1},t_{2}].\ $%
From the equation $\left( \ref{ECsphere}\right) $ we can deduce that
\begin{equation}
\frac{d}{dt}\varphi =cos\theta \
[I_{2}^{-1}(t)-I_{1}^{-1}(t)+(I_{3}^{-1}(t)-I_{2}^{-1}(t))\ sin^{2}\varphi ]
\label{angulos}
\end{equation}%
thus, being $I_{1}(t)<I_{2}(t)<I_{3}(t),$ then $\varphi (t)$ is a \emph{%
monotonous decreasing} function of time, showing that the solution $\Pi (t)$
tends to describe revolutions about the $1$ axis.

\item if \fbox{$E_{max}<\frac{l^{2}}{I_{2_{max}}}$} then $R$ is contained in
either the semi-space $\Pi _{3}>0$ or in $\Pi _{3}<0$. In this case, $\Pi
(t) $ evolves in $S_{L}^{2}$ describing a trajectory that \emph{orbits
surrounding the cartesian axis} $3$, as in the previous case.

\item In \textbf{other cases}, the solution can pass from \emph{orbiting}
one axis to orbit another one. To show this, let us suppose that $%
E_{t_{1}}(\Pi (t_{1}))$ $>\frac{l^{2}}{I_{2_{min}}}$ and that $I_{1}$ is
constant. Then $lenght(\Pi ([t_{1},t_{2}]))\leq l^{2}I_{1}^{-1}(t_{2}-t_{1})$
and so we can choose $I_{1}$ such that $\Pi ([t_{1},t_{2}])$ is contained in
some small patch in $S_{L}^{2}$. In the case that $I_{2}\ $is also constant
in time and $I_{3}(t)$ grows (note that the order is maintained in time),
then $\frac{d}{dt}E_{t}(\Pi (t))=\Pi _{3}^{2}(t)\frac{d}{dt}I_{3}^{-1}(t)<0.$
So the energy decreases as fast as we want if we make $I_{3}(t)$ grow
sufficiently fast. Note that $\Pi _{3}^{2}(t)$ is bounded from below because
$\Pi ([t_{1},t_{2}])$ is in a small patch. In this situation, $E_{t_{2}}(\Pi
(t_{2}))$ can be made smaller than $\frac{l^{2}}{I_{2_{min}}},$ so the
solution is able to pass from the regime $(1)$ to the regime $(2)$ described
above when the energy $E_{t}$ \emph{"crosses"} the homoclinic energy
boundary $\frac{l^{2}}{I_{2min}}.$
\end{enumerate}

\begin{remark}
\emph{(Return time)} In either of the previous cases $(1)$ or $(2)$ we can
give a lower bound for the (shortest) return time $T=t_{2}-$ $t_{1}$ s.t. $%
\Pi (t_{1})=\Pi (t_{2})$. If we suppose that the solution starting at $\Pi
(t_{1})$ satisfies the conditions of $(1)$ above and that it returns to this
value for the first time at $t_{2},$ then%
\begin{equation*}
T=t_{2}-t_{1}\geq \frac{2\pi \left\Vert \Pi (t_{1})\times (1,0,0)\right\Vert
}{l^{2}I_{1_{max}}^{-1}}.
\end{equation*}%
The case corresponding to $(2)$ is analogue.
\end{remark}

Now, suppose that we are in the case considered in $(1)$ ($(2)$ is
analogous) above and that, in some interval $[t_{i},t_{i+1}]\subseteq
\lbrack t_{1},t_{2}]$, the solution $\Pi (t)$ describes a closed simple
curve in $S_{L}^{2}$. Then, we can apply formula $\left( \ref{Mont}\right) $
to find the corresponding phase. Taking into account the time orientation of
the closed solution $\Pi (t)$ (fixed by $\left( \ref{angulos}\right) $), if $%
\frac{area(\tilde{D})}{l^{2}}<2\pi $ (resp. $>2\pi $) we must then take the $%
+$ (resp. $-$) sign in $\left( \ref{Mont}\right) $ and we have that%
\begin{equation}
\pm \frac{area(\tilde{D})}{l^{2}}+\frac{2}{l}\ E_{min}\ (t_{i+1}-t_{i})\leq
\theta _{M}\leq \pm \frac{area(\tilde{D})}{l^{2}}+\frac{2}{l}\ E_{max}\
(t_{i+1}-t_{i}).  \label{bounding tita}
\end{equation}

\begin{remark}
\emph{(Bounding }$\theta _{M}$ \emph{mod. }$2\pi $\emph{) }Note that, since $%
\theta _{M}$ is defined mod. $2\pi ,$ the above bounds yield \emph{%
nontrivial }information when $\frac{2}{l}\ (E_{max}-E_{min})\
(t_{i+1}-t_{i})<2\pi $.
\end{remark}

\subsection{Examples}

We now apply the previous techniques to obtain estimates for the
motion of simple classes of deforming bodies.

\begin{example}
\emph{(Vibrational deformation)} In this case, we suppose that the body is
globally shrinking or expanding, that is, the position of a particle from
the given reference frame $\tilde{S}$ is
\begin{equation*}
r_{i_{0}}(t)=a(t)r_{i_{0}}
\end{equation*}%
where $r_{i_{0}}$ is a constant vector and $a(t)$ is a never vanishing
positive scale factor. This means that we can choose the curve $d_{0}(t)$ in
$Q_{0}$ such that%
\begin{equation*}
I(d_{0}(t))=a^{2}(t)I_{0}
\end{equation*}%
with $I_{0}$ the constant inertia tensor corresponding to the constant
configuration $\{r_{i_{0}}\}.$ By a constant rotation, we can choose the
reference system $\tilde{S}$ (equivalently, another curve $d_{0}(t)$) from
which $I_{0}$ is diagonal. Then, eq. $\left( \ref{ECsphere}\right) $ on the
sphere becomes%
\begin{equation*}
\frac{d}{dt}\Pi =a^{-2}(t)(\Pi \times I_{0}^{-1}\Pi ).
\end{equation*}%
Given an initial value $\Pi (t_{1}=0),$ eq. $\left( \ref{ECsphere}\right) $
can be exactly solved yielding%
\begin{equation*}
\Pi (t)=\Pi _{RB}(\int_{0}^{t}a^{-2}(s)\ ds),
\end{equation*}%
with $\Pi _{RB}$ denoting the rigid body solution of Euler equations $\dot{%
\Pi}=\Pi \times I_{0}^{-1}\Pi $ with initial value $\Pi (t_{1}=0).$ Now, the
function
\begin{equation*}
f(\Pi )=\frac{1}{2}\Pi \cdot I_{0}^{-1}\Pi
\end{equation*}%
is constant along the solutions. Note that $\Pi (t)$ describes a closed
simple curve on the sphere for $t\in \lbrack 0,T]$ when $%
\int_{0}^{T}a^{-2}(t)\ dt=T_{RB}$ equals the period of the rigid body
solution $\Pi _{RB}$. In that case, the corresponding phase is
\begin{eqnarray*}
\theta _{M} &=&-\Lambda _{RB}+\frac{2}{\left\Vert L\right\Vert }\ f(\Pi
)\int_{0}^{T}a^{-2}(t)\ dt \\
&=&-\Lambda _{RB}+\frac{2}{\left\Vert L\right\Vert }\ f(\Pi )T_{RB}
\end{eqnarray*}%
where $\Lambda _{RB}$ is the (signed) solid angle enclosed by the
rigid body periodic solution $\Pi _{RB}$ with energy $f(\Pi )$.
Notice that this phase \emph{coincides with the rigid body phase}
for $\Pi _{RB}$ (\cite{Montph}). The motion of this kind of
vibrating bodies is similar to rigid body motion up to a time
reparameterization which is induced by the expansion/contraction.
\end{example}

\begin{example}
\emph{(Expansion/Contraction of an axially symmetric body)}\textbf{: }Let us
consider the case of an axially symmetric body which expands in the
direction of its symmetry axis, i.e., the case in which there exists a curve
$d_{0}(t)$ such that%
\begin{equation*}
I(d_{0}(t))=diag(I_{1}(t),I_{2},I_{3}),
\end{equation*}%
with $I_{2}=I_{3}.$ As in the previous case, eq. $\left( \ref{ECsphere}%
\right) $ can be exactly solved:%
\begin{equation*}
\Pi (t)=\Pi _{RB}(\int_{0}^{t}\frac{(I_{1}^{-1}(s)-I_{3}^{-1})}{%
I_{1}^{-1}(0)-I_{2}^{-1}}\ ds),
\end{equation*}%
with $\Pi _{RB}$ the rigid body solution to Euler equations $\dot{\Pi}=\Pi
\times I^{-1}(d_{0}(t_{1}=0))\Pi $ and initial value $\Pi (t_{1}=0)$. The
function $f(\Pi )=\frac{1}{2}\Pi \cdot I^{-1}(0)\Pi $ is again constant
along the solution $\Pi (t)$, which is closed simple curve for $t\in \lbrack
0,T]$ when $\int_{0}^{T}\frac{(I_{1}^{-1}(s)-I_{3}^{-1})}{%
I_{1}^{-1}(0)-I_{2}^{-1}}\ ds$ equals the rigid body period $T_{RB}$
corresponding to $\Pi _{RB}$. In that case, the associated phase is%
\begin{equation*}
\theta _{M}=-\Lambda _{RB}+\frac{1}{\left\Vert L\right\Vert }\int_{0}^{T}\Pi
(t)\cdot I^{-1}(d_{0}(t))\ \Pi (t)dt,
\end{equation*}%
where $\Lambda _{RB}$ is the (signed) solid angle enclosed by the rigid body
periodic solution $\Pi _{RB}$ with constant energy $f(\Pi )$. Notice that,
in general, this phase is \emph{different} from the rigid body phase
associated to $\Pi _{RB}$.
\end{example}

\begin{example}
\label{antenna}\emph{(An antenna coming out from a satellite along a
principal axis)}\textbf{: }We now consider the cases in which%
\begin{equation*}
I(d_{0}(t))=diag(I_{1}(t),I_{2_{0}},I_{3_{0}})
\end{equation*}%
or
\begin{equation*}
I(d_{0}(t))=diag(I_{1_{0}},I_{2_{0}},I_{3}(t))
\end{equation*}%
with $I_{1}(t)$ (or $I_{3}(t)$) an increasing function of time. These cases
give simplified models for the situation in which an antenna comes out from
an orbiting satellite along one of the principal axes of inertia $1$ or $3.$
Note that the satellite is free to rotate around its center of mass and so
its motion can be described by eq. $\left( \ref{ecvelpi}\right) $. Suppose
that initially $I_{1}<I_{2}<I_{3}$. Then, in the first case, as $I_{1}(t)$
grows this relation might stop holding after some time, so the solution
could pass from orbiting one axis to orbit another one. Consequently, We
have no control on this kind of solution. More precisely, as
\begin{equation*}
\frac{d}{dt}E_{t}(\Pi (t))=\frac{1}{2}\Pi _{1}^{2}(t)\frac{d}{dt}%
[I_{1}^{-1}(t)]
\end{equation*}%
is negative implying that the energy decreases and the solution can pass
from the case $(1)$ to $(2)$ of the previous section, describing an open
curve on the sphere which we cannot characterize in general. In turn, in the
second case the ordering prevails and%
\begin{equation*}
\frac{d}{dt}E_{t}(\Pi (t))=\frac{1}{2}\Pi _{3}^{2}(t)\frac{d}{dt}%
[I_{3}^{-1}(t)]
\end{equation*}%
is also negative. Since the energy decreases, if the initial value $\Pi
(t_{1})$ corresponds to case $(2)$ of the previous section, the solution
also evolves according to $(2)$ and we have a good characterization of its
behavior. In particular, if $\Pi ([t_{i},t_{i+1}])$ is a closed simple
curve, we then know that the corresponding reconstructed rotation $%
R(t_{i+1}) $ is $exp(\theta _{M}\ \frac{\hat{L}}{\left\Vert L\right\Vert })\
R(t_{i})$ where from $\left( \ref{bounding tita}\right) $,%
\begin{equation*}
\pm \frac{area(\tilde{D})}{\left\Vert L\right\Vert ^{2}}+\frac{2}{\left\Vert
L\right\Vert }\ E_{min}\ (t_{i+1}-t_{i})\leq \theta _{M}\leq \pm \frac{area(%
\tilde{D})}{\left\Vert L\right\Vert ^{2}}+\frac{2}{\left\Vert L\right\Vert }%
\ E_{initial}\ (t_{i+1}-t_{i}),
\end{equation*}%
with $E_{initial}$ the initial (hence the maximum) value of the energy $%
E_{t} $ in $[t_{i},t_{i+1}]$ and $E_{min}=E_{t_{i+1}}$. We thus note that we
can have a better description of the motion of the satellite when the \emph{%
antenna comes out along the largest principal axis of inertia}.
\end{example}

\begin{remark}
\emph{(Slow deformations)\ }Intuitively, when the antenna comes out \emph{%
very slowly}, the motion of the satellite will be close to a rigid body
motion. This is reflected in the fact that when $\frac{d}{dt}[I_{3}^{-1}(t)]$
is \emph{very small} (compared to $\frac{l^{2}E_{initial}}{(t_{i+1}-t_{i})}$%
), then $E_{min}\sim E_{initial}$ and $\frac{area(\tilde{D})}{\left\Vert
L\right\Vert ^{2}}\sim \Lambda _{RB}$. So the phase $\theta _{M}$ is \emph{%
approximately the same as the rigid body phase} associated to a rigid body
with $I=I(d_{0}(t_{i}))$ and initial $\Pi (t_{i})$.
\end{remark}

\begin{remark}
\emph{(Small bodies in the gravitational field)} For a body, v.g. a
satellite orbiting the earth, which is \emph{small} with respect to the
interaction distance with another body (eg: the earth), it is a very good
approximation to suppose that the gravitational force acting on a particle
of mass $m_{i}$ of this body is%
\begin{equation*}
F_{i}=m_{i}\ \frac{-G\ M}{(r_{CM}-P)^{3}}(r_{CM}-P)
\end{equation*}%
where $r_{CM}$ denotes the position of the center of mass of the body. $P$
and $M$ denote the position of the center of mass and the total mass of the
second body (eg: the earth), respectively. We thus see that the equations of
motion for the position of the center of mass (a \emph{central force problem}%
) are totally decoupled from the equations of motion giving the rotation
about the center of mass (a \emph{self deforming body problem} as in ex. \ref%
{antenna}).
\end{remark}

\end{document}